# A New GEM-like Imaging Detector with Electrodes Coated with Resistive Layers

A. Di Mauro, B. Lund-Jensen, P. Martinengo, E. Nappi, V. Peskov, L. Periale, P.Picchi, F. Pietropaolo, I.Rodionov

*Abstract*--We have developed and tested several prototypes of GEM-like detectors with electrodes coated with resistive layers: CuO or CrO. These detectors can operate stably at gains close to $10^5$ and they are very robust. We discovered that the cathodes of these detectors could be coated by CsI layers and in such a way the detectors gain high efficiency for the UV photons. We also demonstrated that such detectors can operate stably in the cascade mode and high overall gains (~$10^6$) are reachable. This opens applications in several areas, for example in RICH or in noble liquid TPCs. Results from the first applications of these devices for UV photon detection at room and cryogenic temperatures are given.

## I. INTRODUCTION

The GEM detector has several unique features, for example, it can operate at rather high gains in pure noble gases, can be combined with another GEMs and operate in cascade mode and so on [1].
However, the GEM is a rather fragile detector: it requires dust free conditions for its assembling and could be easily damaged by sparks. Unfortunately, occasional sparks are almost unavoidable at high gains of operation. Several groups tried to minimize their rate and damaging effect by using segmented GEMs [2], or many GEMs (up to 4 -5) in cascaded mode [3] or by identifying the optimal combination of the parameters (the width of the gaps between the cascaded GEMs, their voltages at a given counting rate and so), which ensure the minimum rate of sparks at the given overall gain and by respecting this "safe zone" during the operation [4].
Our group also tried to contribute to these efforts.
Performed studies show that the maximum achievable gain of hole- type detectors increases with their thickness [5]. This is why our first attempt was to develop a thick GEM (TGEM) [6, 7]. This was a metallized from both sides printed circuit board with drilled holes-see Fig. 1. This simple device allows one to achieve the maximum gain 10 times higher than with the conventional GEM [7]. Later we modified this detector by drilling out a Cu layer around each hole; this allowed one to additionally increase the maximum achievable gain by a factor of ~5. A systematic study of this device was performed by Breskin's group: they confirmed the detector is very robust and can operate at gains of ~$10^5$. Instead of drilling out the Cu around the edges of the holes, they manufactured the protective dielectric rims by a lithographic technology [8].

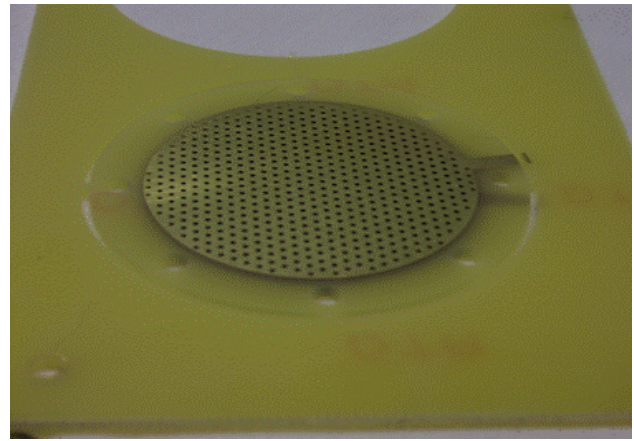

Fig.1. A photograph of a TGEM.

Recently we have developed and tested a thick GEM which electrodes were coated by a thick layer of graphite paint [9]. We named this detector a Resistive Electrode Thick GEM or RETGEM. The RETGEMT could operate at gains of ~$10^5$; at higher gains it may transit to a streamer mode and continue to operate in this mode as a photon counter. In contrast to sparks in conventional GEMs these streamers are mild (see Fig. 2) and do not damage either the detector or the front-end electronics.

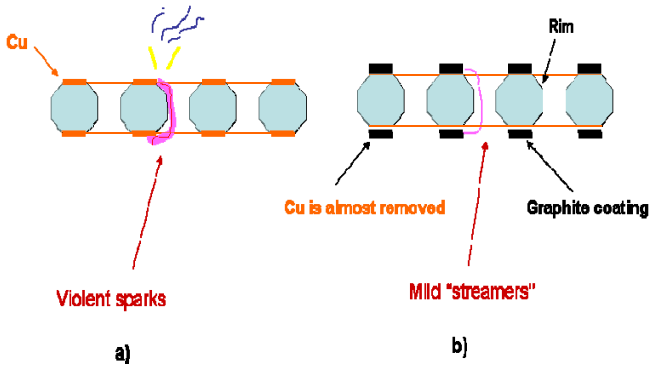

Fig.2. A schematic drawing showing the difference between the sparks and 'streamers": in the case of the spark occurring in the GEM detector (a) all electrical energy stored in the detector's capacity released in the spark; in the case of the restive coating the discharge current is restricted by the electrode's resistivity and by its charging up effect and as a results the discharges (streamers) are very mild (b).

Certainly there is nothing magic in the graphite coating and many other resistive layers could be used to achieve the same protective effect. In this work we experimented with TGEMs coated with thin oxides layers: CuO or CrO. Due to their small thickness of these coatings they do not restrict the spark's energy as much as the graphite layer used in [9], but nevertheless they made the detector very robust.
The aim of this work is to extensively test these detectors and use them for UV photon detection at room and cryogenic temperatures.

## II. RETGEMs WITH CuO OR CrO COATED ELECTRODES

As in our previous work [9] RETGEMs for these studies were manufactured by coating TGEMs electrodes with resistive layers. The last one were produced from G-10 sheets (3x3, 5x5 or 10x10cm$^2$) using the industrial PCB processing of precise drilling and etching. The TGEM used were 0.4 -1,5 mm thick with holes of 0.3 -1mm in diameter and with a pitch of 0.7-2.5 mm, respectively. Their electrodes were made of Cu or Cr and in all detectors the electrodes were etched around the hole edges in order to remove sharp edges and create dielectric rims of 0.1-0.15 mm in width. For the sake of simplicity for this work, the resistive coating was done by oxidation of the metallic electrodes. The photograph of the one of the prototypes of these detectors is presented in Fig. 3

Note that these detectors were very different from our first prototypes, described in [9], in which the Cu electrodes were etched until they become very thin and nonuniform in structure, then they were coated by thick graphite layers and the edges of the hole were then additionally drilled out to remove sharp edges. Thus it was not clear in advance if these new designs (with CuO or CrO layers) would provide any spark protection.
The experimental sent up for their study is shown in Fig.4. It contains two gas chambers connected together by a pipe line and flushed by the same gas at a pressure of 1 atm. In one of the chambers a RETGEM was installed and in another one, a GEM, which we used for comparative studies. Most of the GEMs used in these studies were had sizes of 10x10cm$^2$ and were manufactured at CERN. However in some studies we also used GEMs manufactured in the USA [10]. As ionization sources we used 55Fe or $^{241}$Am radioactive sources placed inside the chambers. Signals from the detectors were recorded by charge sensitive amplifiers Ortec142A or CANBERRA.

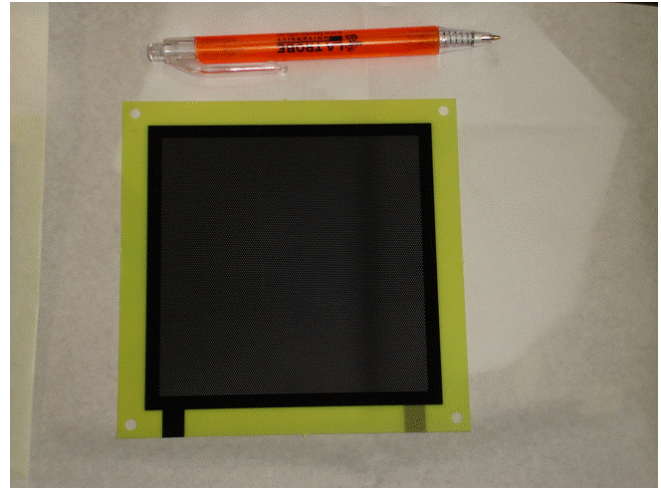

Fig. 3. A photo of the larger (10x10cm$^2$) prototype of the RETGEM with CuO electrodes.

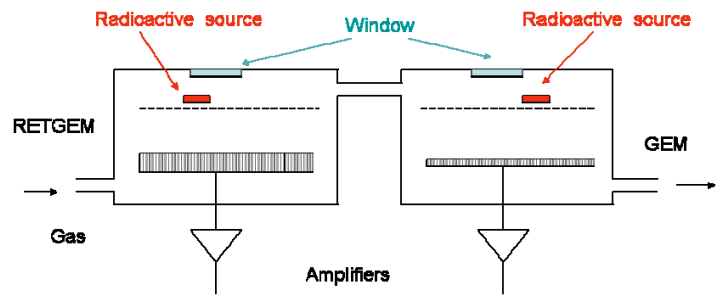

Fig. 4 A schematic drawing of the experimental set up for comparative studies of the RETGEMs and the GEMs

Some results of gain vs. voltage measurements are presented in Fig. 5. The measurements were stopped at voltages at which the first signs of gain instability appeared. One can see from this data that the RETGEM operates stably in Ar at gains of 10 times higher than the GEM. At gains close to $10^5$ discharges may appear in the RETGEM. Because the oxide layers were much thinner than the graphite coating we used in the earlier studies [9] the discharges in the present version of the RETGEM were not mild streamers, but rather sparks. However, the energy released in these sparks was less then in the case of the TGEMs and as a result the detector was more robust than the TGEM

Fig. 6 shows a typical oscillogramm of the signal from the RETGEM operating in Ar at gains close to $10^4$. One can clearly see the fast component of the charge signal induced by avalanche electrons and the slow component due to the drift of positive ions. It is interesting to see that at gains of close to breakdown, no signs of ion feedback pulses were observed indicating that the breakdown occurs via the space charge mechanism [5].

Because RETGEMs operate at gains much higher than GEMs, it was attractive to use them for single electron detection and as photodetectors. First results obtained in this direction are presented in the next paragraph.

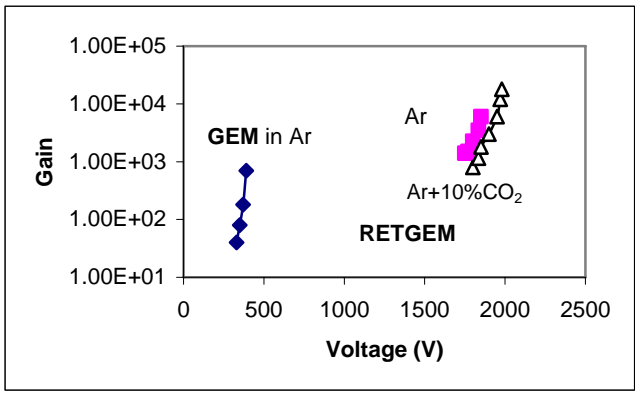

Fig. 5. Gains vs. voltage measured for the GEM operating in Ar and for the RETGEM (1 mm thick) in Ar and Ar+10%$CO_2$.

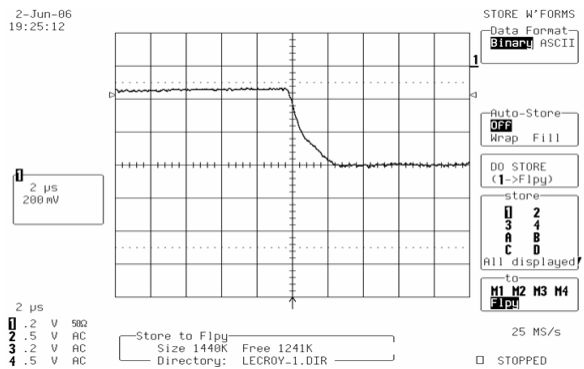

Fig. 6. Typical oscillogramm of the signal produced by 5.9 keV photons in a 1 mm thick RETGEM at gain of $5\times 10^3$.

## III. TESTS OF RETGEM –BASED PHOTODETECTORS

Several groups (see for example [11, 12] demonstrated that cascaded GEMs (3-4 GEMs operating in tandem) combined with semitransparent or reflective CsI photocathodes could be used for the detection of UV and even visible photons [12]. This detector's configuration offers new possibilities in some applications, for example in the detection of the Cherenkov light. Indeed GEMs with reflective photocathodes can operate and remain high sensitive to light at zero or even at the reverse drift electric field being in such a way a "hadron blind" (see [13] for more details). Moreover, in some cases GEMs can be placed and be operated in the same gas as a Cherenkov radiator so that no separation windows are needed between them.

Thus is will be interesting to evaluate if the RETGEM can offer a comparable or even better performance.

Because the RETGEM has a dielectric coating it is not clear in advance if it could be coated with an CsI film or any other photosensitive layers and if these layers remain stable and have a high enough quantum efficiency. It was not evident as well that these detectors can operate stably in cascaded mode. To answer these questions and investigate other potential problems we build prototypes of cascaded RETGEMs combined with CsI or SbSc photocathodes and performed their preliminary tests.

### A. Tests oriented on RICH applications

For these tests we slightly modified the experimental set up shown in Fig. 4. Inside the first chamber two RETGEMs operating in cascade mode were installed (we named them 'double RETGEMs" and inside the other one –three cascaded GEMs ("triple" GEMs) with Cu electrodes manufactured by Tech-Etch Inc. [10]-see Fig. 7. The cathode of the first (top) RETGEM and the Cu cathode of the top GEM was coated with an CsI layer 0.4 mm thick (by a vacuum deposition technique). From our earlier experience we know that the Cu substrate may cause a rather fast degradation of the CsI quantum efficiency (QE), this is why it was very important not only to measure the initial value of the QE immediately after the CsI evaporation but also monitor it in time. This was done with the help of a Hg lamp. The UV light from the Hg lamp entered the detectors via the $CaF_2$ windows covered with narrow band filters having a peak transmission at 185 nm. By applying the negative (reverse) voltage between the top GEM electrode with the CsI photocathode and the drift mesh, the photocurrent was measured in various gases as a function of this voltage. In the case of the mixtures of noble gases with quenchers, this photocurrent reaches a plateau at high voltages with a value of $I_{GEM}$ (see Fig. 8) which could be interpreted as "full" collection of the photoelectron from the photocathode:

$$I_{GEM} = kI_{GEMvac} \quad (1),$$

where $I_{GEMvac}$ is the current from the GEM photocathode measured in vacuum and the $k$ is the coefficient describing a back diffusion effect [14]. Usually in quenched gases the value of k is close to one, whereas in noble gases $k<1$.

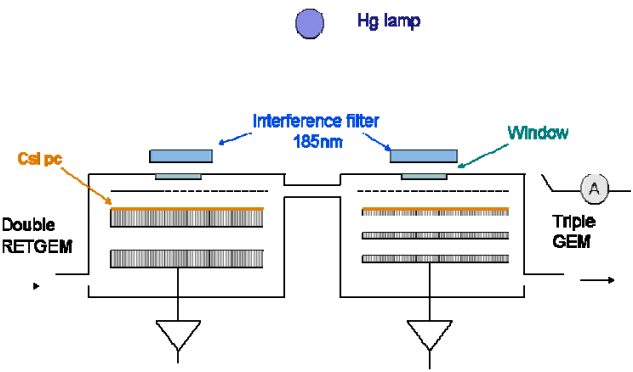

Fig. 7. A schematic drawing of the experimental set up used for comparative studies of RETGEMs and GEMs

To evaluate from these data the GEM's QE we used photodiodes R1259 and R1187 calibrated at Hamamatsu. The photocurrent from these photodiodes exhibited a very clear plateau (with a value at the plateau $I_{PD}$) and by comparison the values of these photocurrents and taking into account the solid angles at which the UV light reached the detectors, one can calculate the QE of the GEM being 13.3%. Of course, for the evaluation of the GEM's quantum efficiency operating in the counting mode ($Q_{pract}$) one have to take into consideration the photoelectron collection factor ε (see [3] for details) which could be obtained for example, from the measurements of current $I_A$ from the anode of the bottom GEM($ε=I_{GEM}/I_A$). However, in our measurements we observed that the $I_A$ steadily increased with the applied voltages (no clear plateau was observed) and thus these simple measurements did not provide any reliable data for the calculation of the $Q_{pract}$. Obviously, the measurements should be performed in counting mode as was done in [3] and this will be our future task. However, coming from the results presented in [3], one can expect that ε~1 at overall gains of triple GEM ~$10^4$.

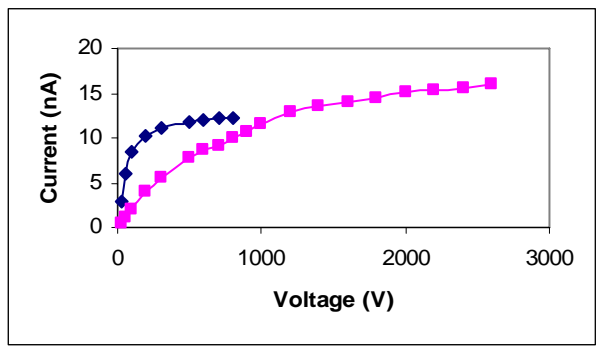

Fig. 8. Current measured from a photodiode R1259 and from the CsI cathode of the top GEM. Note that at high voltages both currents reached plateau: $I_{PD}$= 12.2 nA and $I_{GEM}$=15.9 nA.

We also tried to perform the same current measurements in the case of the RETGEM. Unfortunately, a rather strong charging up effect was observed, even at small values of the photocurrent, so we did not consider these measurements to be reliable for further interpretation. To compare the practical quantum efficiency of the GEM and the RETGEM we performed measurements in a counting mode. For this the UV light from the Hg lamp was very strongly attenuated and we measured under the identical conditions the counting rates from the GEM ($n_{GEM}$) and the RETGEM ($n_{RETGEM}$). For the same overall gains of $10^4$ and the same electronics threshold the ratio of the counting was $n_{RETGEM}/n_{GEM}$= 1, 73. If one assumes that ε~1 even in the case of RETGEM, than the estimated QE for RETGEM will be Qpract~23%. Of course in the nearest feature we will measure the value of ε and this will allow us to estimate Qpract more accurately. However, in this first stage of the work it was important just to have a rough estimate of the Qpract order to be sure it has a reasonably high value even in the case of the CuO substrate and to monitor the photocathode's stability in time. The last one was done by regular measurements of counting rates from the GEM and the RETGEM under identical conditions over a period of three months. No big changes in the counting rates were observed (the variations were on the level of 10%only) either for the GEM or the RETGEM indicating that the CsI photocathodes remained stable for both detectors.

We also performed comparative measurements of maximum gains achievable with double RETGEMs and triple GEMs both coated with the CsI layers. Some results are presented in Figs 9 and 10. The measurements were stopped at voltages when first signs of discharges appeared. One can see from this data that in the case of Ne and Ar, double RETGEMs offer much higher gains than triple GEMs. This feature makes the RETGEM very attractive for RICH applications.

The next set of experiments were performed in order to investigate if another photocathodes (for example one that is sensitive to visible light) could be deposited on the top of the CuO substrate and if it could remain stable afterwards.

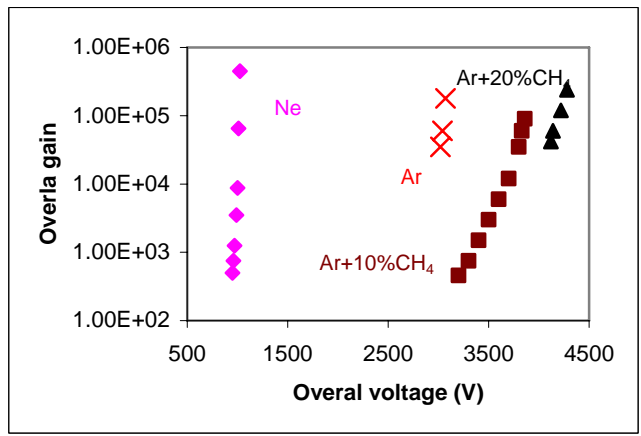

Fig. 9. Gain vs. voltage for double RETGEM measured in various gases

The manufacturing of high quality photocathode sensitive to visible light is a quite complicated procedure [15, 16].

However, some low efficiency photocathodes could be produces in a rather simple way by coating the selected substrate by Cs release from the "Cs generator" [17] in a vacuum of $10^{-6}$Torr. In this work, we used this simple technology. One of the surfaces of the RETGEM (size of 3x3 cm$^2$, 1.5 mm thick) was coated by an Sb layer 0,2 μm thick by a vacuum deposition technique. The RETGEM was then extracted from the evaporation set up and placed inside a quartz tube (the inner diameter of which was 70 mm) and which had several electrical feedthroughs in its metallic flanges-see Fig. 11. The tube with the RETGEM was heated to ~100°C and pumped to a vacuum of $10^{-6}$ Torr for several days. It was then cooled down to room temperature and the Cs generator was activated; Cs vapour released from the generator reacted with the Sb surface and finally formed SbCs.

The main problem associated with this primitive technique is the excess of Cs remaining on the inner walls of the tube and on the surfaces of feedthroughs. Sometime there were cases of current instabilities during the measurements. However, we succeeded to move the Cs depositions out from the chamber into the pumping system by local heating of the contaminated parts of the tube by a small flame. After such cleansing procedures we were able to perform measurements of the photocurrent produced by a lamp and monitor the stability of the photocathode with time. Some our first results are presented in Fig. 12. One can see that immediately after the photocathode's manufacturing, the photocurrent dropped very steadily, but then "stabilized" and began to degrade quiet slowly so that we had enough time for the measurements the photocathode's QE to take place. For this the tube was attached to the monochromator (combined with a Hg or H$_2$ lamp) and the photocurrent Id (λ) produced in the detector by the light from the monhromator was measured as a function of the wavelength. After these measurements were completed, the quartz tube was replaced by a Hamamatsu calibrated photodiode R414 and for each wavelength we measured the photocurrent from the photodiode I$_{PD}$(λ) produced by the light beam exiting the monochromator. From the known absolute QE of the photodiode R414 and the ratio I$_d$(λ)/ I$_{PD}$(λ) the QE of the RETGEM was calculated. Some our fist preliminary results are shown in Fig. 13. One can see that the quantum efficiency achieved by such a manufacturing technique was 2-3 times lower than in the case of the high quality photocathodes, however we consider these preliminary results as rather encouraging, because we believed that in the future tests we will be able to protect the SbCs photocathodes deposited on the to of the RETGEMs by a thin (~20 nm) CsI layer.

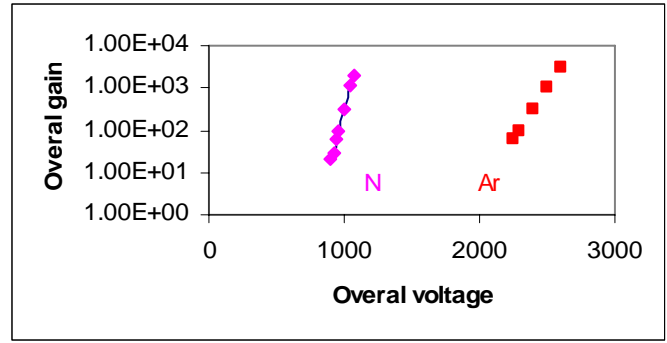

Fig.10. Gain vs. voltage for triple GEM measured in Ne and Ar.

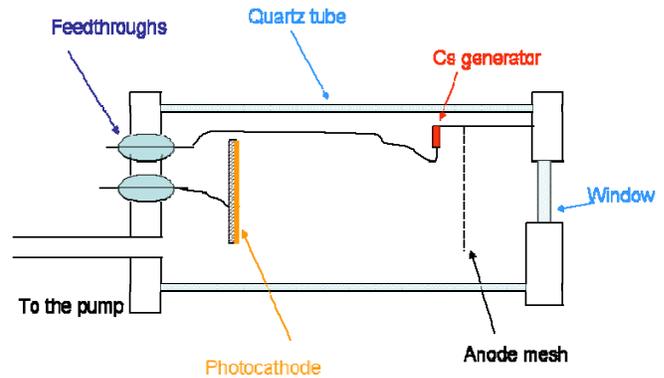

Fig. 11. A schematic drawing of the set up used for manufacturing SbCs photocathode on the top of the RETGEM

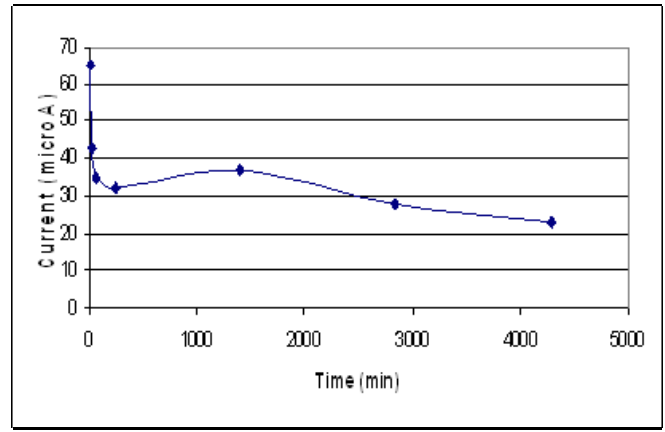

Fig.12. Photocurrent vs. time measured in vacuum between the cathode and the anode mesh after manufacturing the SbCs photocathode

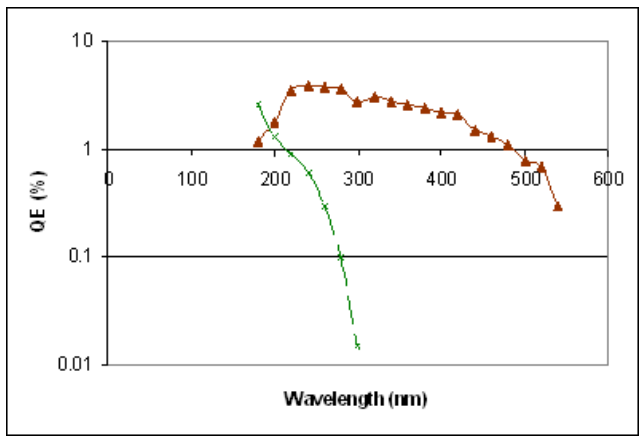

Fig.13. Results of the QE measurements: brown curve- SbCs photocathode, green curve-SbCs photocathode covered by a CsI protective layer

This technique was first described in [17] and later it was further developed by Breskin group (see for example [18] and reference therein).
For the time being however, we coated the RETGEM by a very thick (100 nm) CsI layer using a conventional evaporation set up and the RETGEM was then extracted in air and placed inside the quartz tube which was then immediately evacuated.
The results of the measurements the QE of such photocathode exposed for a few minutes to air are presents in Fig. 13. This photocathode did not show any signs of degradation during one week monitoring of its QE under the vacuum. Certainly, more tests are needed to demonstrate that RETGEMs coated with SbCs or SbCs/CsI photocathodes could stably operate in gas conditions.

*B. Tests oriented on applications for noble liquid TPCs*

In our recent work [19] we have demonstrated that TGEMs coated with CsI photocathodes can operate at cryogenic temperatures and detect the scintillation light from noble liquids (see also [20]).
It will be interesting to check if RETGEMs, in spite of their resistive electrodes, can also operate stably at cryogenic temperatures especially in the case when they are coated by an CsI layer. To verify this, we have performed several sets of measurements with single double RETGEMs cooled to cryogenic temperatures.
Our experiment set up was the same as in work [20] and it is shown schematically in Fig. 14. It consists of the cryostat with a test chamber placed inside it. Depending on the measurements either a single or a double RETGEMs (1mm or 1,5mm) thick) with the top electrode coated by an CsI layer was installed inside the chamber (see Fig. 14) as well as a radioactive source $^{55}$Fe for gas gain measurements. In some experiments a scintillation chamber (see [20] for more details) was attached to the test chamber; it contained an $^{241}$Am alpha source inside was flushed by Ar at a pressure of 1atm. Figs. 15-17 show gain vs. voltage curves measured at room temperature and 100K for RETGEMs 1 and 1,5 mm thick respectively. One can see that gains of $10^4$ could be achieved at 100K with double RETGEMs. Because of our test chamber was flushed with Ar we could, if necessary, liquefy Ar inside the chamber and investigate the operation of the RETGEM in the case when the LAr level was just 1-2 cm below the anode of the RETGEM (see Fig. 14). Results of gain measurements in this condition are shown in Fig. 16. One can see that compared to the case where the RETGEM operated in Ar at 100K, the operating voltage of the RETGEM placed 1-2 cm above the LAr level was higher, indicating that probably a thin layer of LAr was formed on the surface of the RETGEM.

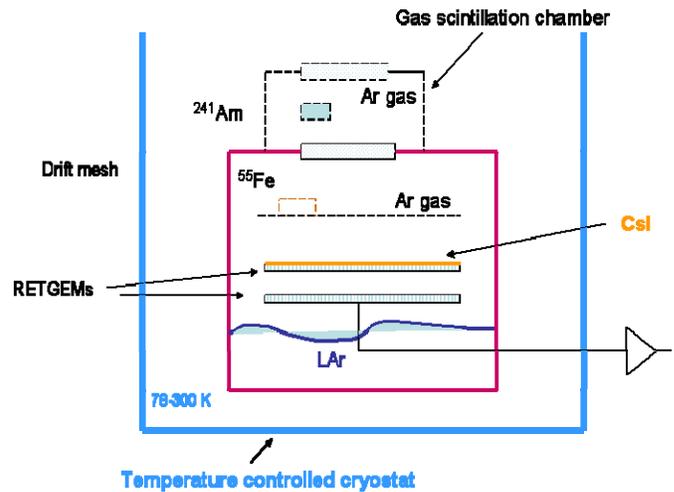

Fig. 14. A schematic drawing of the set up used for measurements with RETGEMs at cryogenic temperatures.

The $Q_{pract}$ of the CsI photocathode at various temperatures can be estimated from the amplitude of the signal B from the

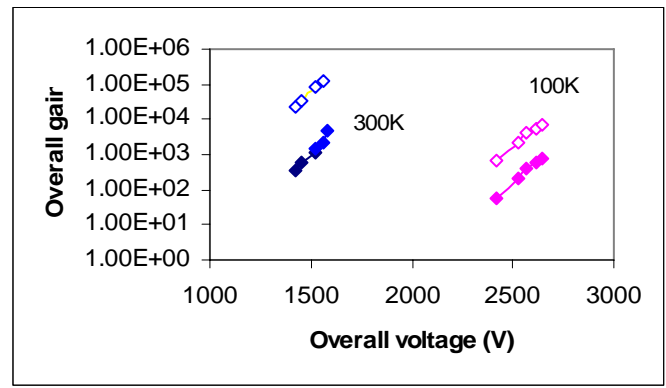

Fig.15. Gain vs. voltage for a single (filled symbols) and for a double (open symbols) RETGEM (1mm thick) measured at room temperature and at 100 K in Ar at pressure of 1atm.

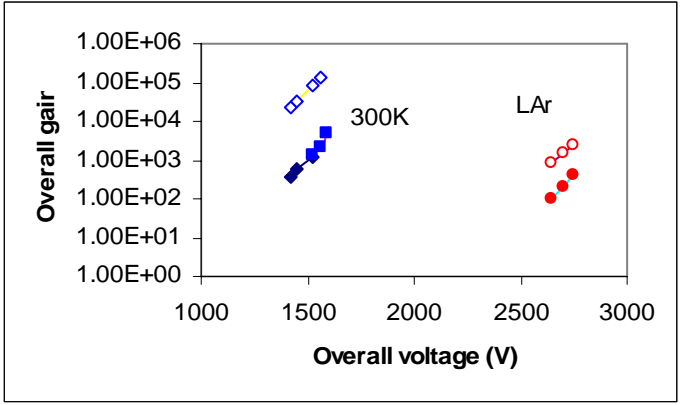

Fig. 16. Gain vs. voltage for a single (filled symbols) and for a double (open symbols) RETGEM (1 mm thick) measured at room temperature and in the case when the RETGEM was 1-2 cm above LAr level.

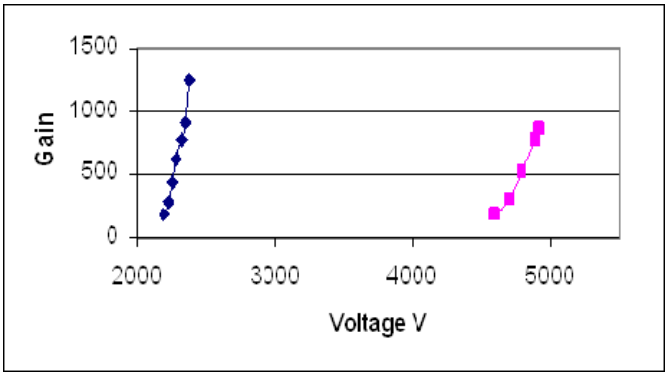

Fig. 17. Gain vs. voltage for a RETGEM (1,5 mm thick) measured in Ar at 1atm at room temperature (blue curve) and 100K (rose curve).

RETGEM detecting the scintillation light produced by alpha particles:

$$B = A N_{ph} \Omega Q_{pract} \quad (2),$$

where A is a gas gain, $N_{ph}$ is the number of UV photons emitted by the alpha source, $\Omega$ is a solid angle at which the scintillation light reaches the CsI cathode.
As was discussed earlier for the RETGEM

$$Q_{pract} = \varepsilon Q k \quad (3),$$

where Q is the QE measured in vacuum (k<1 for Ar). Assuming that

$$N_{ph} = E/W \quad (4),$$

(E is the energy of alpha particles and W is the energy required to produce a UV photon) and assuming that $N_{ph}$ is independent of the gas density, the calculated $Q_{pract}$ was then 28% and 17% at room temperature and 100K, respectively.
These very preliminary results demonstrate that RETGEMs could be an attractive alternative to PMTs or any other type of photodetectors for noble liquid TPCs.

IV. NEW DEVELOPMENTS

Certainly that RETGEMs described in this work and in the previous one [9] are just first prototypes and are far from being ideal. For example, one of the problems of the RETGEMs with the CuO or CrO layers is that at high gains they transit to sparks rather than to mild streamers. The reason for this is quite clear: the key parameter for the spark quenching is the amount of the surface charge from the incoming avalanche that is needed to substantially modify the field in the detector hole, diminishing the energy of the discharge, or in other words, the capacity per unit area. This will be high if the layer is thin or if the dielectric constant is high. This is why in order to strongly reduce the energy of the sparks one has to use rather thick resistive electrodes. One of our successful prototypes is shown in Fig. 18. It is a RETGEM, the top electrode of which (the cathode), is coated by the CuO layer and the bottom (the anode) is a 1-2 mm thick resistive plate made of Pestov glass or any other high resistivity material. This detector at gains $>10^4$ operated in a proportional mode and at gains of $>10^5$ transited to a streamer mode. We are now investigating several other designs based on this principle.

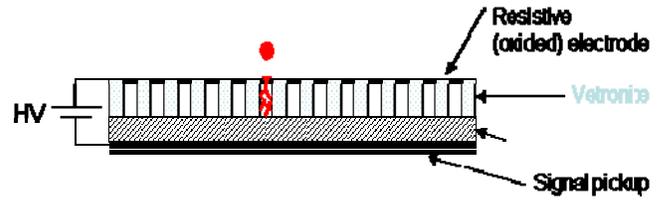

Fig. 18. A schematic drawing of the RETGEM the top electrode of which was coated by a CuO layer and the bottom electrode (the anode) was a thick high resistivity plate.

V. DISCUSSION AND CONCLUSIONS

The obtained preliminary results demonstrate the potentials of the new detector. In spite of the fact that at high gains it transits to sparks rather than to a streamer, it is much more robust than the GEM or even the TEGEM. The other important discovery was that the RETGEM could be combined with photocathodes and can operate in cascade mode.
We believe that the suggested detectors after some improvements will open new possibilities for applications which do not require extremely high counting rates or very good position resolutions, for example in RICH, cryogenic TPCs or UV visualisation in daylight conditions [9].